# RADIATIVE HABITABLE ZONES IN MARTIAN POLAR ENVIRONMENTS

Carmen Córdoba-Jabonero[1], María-Paz Zorzano[1], Franck Selsis[1], Manish R. Patel[2] and Charles S. Cockell[3]

[1]*Centro de Astrobiología (CSIC-INTA), Ctra. Ajalvir km. 4, Torrejón de Ardoz, 28850-Madrid (Spain)*

[2]*Planetary and Space Sciences Research Institute, The Open University, Walton Hall, Milton Keynes MK7 6AA (UK)*

[3]*British Antarctic Survey, High Cross, Madingley Road, Cambridge CB3 0ET (UK)*

Number of Pages: 44

Number of Figures: 8

Number of Tables: 1

*Proposed running head (not more than 55 characters):*

Radiative Habitable Zones at the Martian Poles

Current corresponding author's mailing address:

Carmen Córdoba-Jabonero, PhD

Instituto Nacional de Técnica Aeroespacial (INTA)

Dpt. Observación de la Tierra, Teledetección y Atmósferas

Área de Investigación e Instrumentación Atmosférica

Ctra. Ajalvir km. 4, Torrejón de Ardoz, 28850-Madrid, Spain

Phone: +34 915202009

Fax: +34 915201317

e-mail: cordobajc@inta.es

**Abstract**

The biologically damaging solar ultraviolet (UV) radiation (quantified by the DNA-weighted dose) reaches the Martian surface in extremely high levels. Searching for potentially habitable UV-protected environments on Mars, we considered the polar ice caps that consist of a seasonally varying $CO_2$ ice cover and a permanent $H_2O$ ice layer. It was found that, though the $CO_2$ ice is insufficient by itself to screen the UV radiation, at $\sim$ 1 m depth within the perennial $H_2O$ ice the DNA-weighted dose is reduced to terrestrial levels. This depth depends strongly on the optical properties of the $H_2O$ ice layers (for instance snow-like layers). The Earth-like DNA-weighted dose and *Photosynthetically Active Radiation* (PAR) requirements were used to define the upper and lower limits of the northern and southern polar radiative habitable zone (RHZ) for which a temporal and spatial mapping was performed. Based on these studies we conclude that photosynthetic life might be possible within the ice layers of the polar regions. The thickness varies along each Martian polar spring and summer between $\sim$ 1.5 m and 2.4 m for $H_2O$ ice-like layers, and a few centimeters for snow-like covers. These Martian Earth-like radiative habitable environments may be primary targets for future Martian astrobiological missions. Special attention should be paid to planetary protection, since the polar RHZ may also be subject to terrestrial contamination by probes.

**Introduction**

At the surface of Mars, ultraviolet (UV) radiation levels at wavelengths shorter than ~290 nm, the most biologically harmful UV radiation, are much higher than those encountered on Earth. Solar UV radiation at wavelengths less than 200 nm is totally removed due to the high absorption of $CO_2$ (a major component of the Martian atmosphere), but longer wavelength UV radiation (200 - 400 nm) does reach the surface. These high UV radiation fluxes have been recognized as a potential problem for any putative biota (Sagan and Pollack, 1974; Kuhn and Atreya, 1979). More recently, the surface UV flux has been investigated using different radiative transfer models, which produce similar results (Cockell *et al.* 2000; Patel *et al.* 2002; Córdoba-Jabonero *et al.* 2003). The sterilizing effects of UV radiation under simulated Martian conditions have also been experimentally demonstrated (Schuerger *et al.* 2003).

All known life forms on Earth share a common feature: their genetic information is coded in a DNA or RNA chain of nucleotides. When exposed to sufficiently high levels of UV radiation these chains are damaged. Therefore, organisms must either have UV protection mechanisms or efficient repair processes. Some possible atmospheric means of protection have been proposed (Córdoba-Jabonero *et al.* 2003; Patel *et al.* 2004), such as atmospheric dust/aerosol scattering and $SO_2$ of volcanic origin at some point of the history of Mars. However, on present-day Mars there is no large scale active volcanism. Furthermore, the UV shielding by dust scattering, in spite of its high content, does not screen a particular wavelength range as in the case of molecular absorption, allowing all wavelengths of the solar UV spectrum greater than 200 nm to reach the surface with an attenuated intensity.

Therefore, the existence of life on Mars, at least at the surface, cannot be considered as probable. The first discussion on the existence of protected environments suitable for photosynthetic organisms on Mars was treated in the work of Sagan and Pollack (1974). These authors considered the UV shielding together with visible (VIS) radiation penetration in Martian soils and concluded that a favorable zone for photosynthetic could exist, although they did not deal with polar ices directly. Searching for potentially habitable environments, and following previous investigations of UV protection in terrestrial ice-covered microhabitats (Cockell *et al.* 2002; Cockell and Córdoba-Jabonero 2004), we studied the habitability of the Martian subsurface polar environment. Such regions are excellent candidates to shelter biological matter due to two special conditions: 1) they are at a high latitude (surface radiation levels are much lower due to higher solar zenith angles, and possible $O_3$ partial protection in winter/spring), and 2) they are ice covered, providing an important biological protection due to the high absorption of UV radiation by $CO_2$ ice (Hansen, 1997) and $H_2O$ ice (Warren, 1984; Perovich, 1993).

Life, as we know it on present-day Earth, has adapted to a range of physical extremes, defining thus the habitability of an environment. From the radiative (UV and VIS) point of view, one limiting factor is the biological sensitivity to the UV spectrum reaching the surface/subsurface. In addition to this radiative condition, photosynthetic organisms require exposure to solar radiation since they use the VIS region of the spectrum as a source of energy and are therefore also exposed to damaging UV radiation. Therefore, Martian polar zones fulfilling both conditions are eligible radiative habitable zones (RHZ) for Earth-like photosynthetic organisms.

There are many other considerations important for the habitability of an environment. One requirement for life is the existence of liquid water. The polar regions are good candidates for the existence of liquid water because of the presence of abundant frozen water near the surface. According to a recent study on Mars the heating must be sudden for the $H_2O$ ice to melt into a liquid state before it sublimates, a condition that is satisfied by locations on almost perpetual winter darkness, i.e. near the poles (Hecht 2002), where there is a strong change of the radiative-heating conditions from winter to spring season. Furthermore, the solid state greenhouse effect may warm the region below the ice (Matson and Brown 1999) allowing for temperature conditions more favorable for life and for the existence of liquid water. Adsorbed liquid water and thin films of liquid water can exist at subzero temperatures when grains of ice are in contact with each other or with individual soil grains (Jakosky *et al.* 2003). These aqueous veins in the ice permit the transport of nutrients to and products from microbes (Price and Sowers 2004). Finally the requirement that surface pressures exceed 6.11 mbar (the triple point pressure) for pure liquid water to exist is satisfied in most of the northern hemisphere poleward of 30°N for virtually the entire year (Haberle *et al.* 2001). Therefore the presence of liquid water in the north polar region, if only in minimal quantities, may be speculated as possible although its existence has not been directly proved.

The low temperatures of these ices may be a limiting factor but several authors have shown that microbial metabolism can occur below zero (Rivkina *et al.* 2000; Jakosky *et al.* 2003; Price and Sowers 2004). The surface temperatures of the ice polar caps range between -130°C at mid-winter and -20°C at mid-summer. These temperatures may have been higher in the past, for different Martian obliquities (Costard *et al.* 2002; Jakosky *et al.* 2003). Recent experiments with microbial communities report that metabolism is active even at -40°C (Price and Sowers 2004). A terrestrial ecosystem that may serve as analogue is found in the surface snow of the interior of the Antarctic continent, where temperatures range between -85°C and -13°C and still show large populations of bacteria with active metabolism down to -17°C (Carpenter *et al.* 2000). Other comparable and widely studied, terrestrial ecosystems are snow algae (Hoham *et al.* 1983).

Finally there are other harmful radiation sources reaching Mars: ionizing and neutron radiation caused by galactic cosmic radiation and solar particle events. Due to the lack of a magnetic field and the low shielding of the Martian atmosphere (the Martian overhead airmass is 16 g cm$^{-2}$ instead of the terrestrial 1000 g cm$^{-2}$) the doses of ionizing radiation at the surface of Mars reach values about 100 times higher than those on the Earth (Baumstark-Khan and Facius 2002). However, since a great variety of microbes tolerate this type of radiation at similar or even greater doses than those found on Mars, ionizing radiation cannot be considered a limiting factor for microbial life on Mars (Baumstark-Khan and Facius 2002; Gilichinsky 2002) and thus here we will limit our study to solar UV shielding and VIS radiation penetration.

In summary, on the surface of Mars, UV radiation is the key problem for life on Mars. In this work the polar regions of Mars were considered as candidates to shelter life because of the abundance of $H_2O$ near the surface, if only in the frozen state, and because of the UV screening properties of ices. The radiative study is explained in detail for the north pole case, which in terms of pressure-temperature (liquid water availability) and surface UV levels is a more favorable environment for life. The results of the equivalent analysis for the south pole, which is a more extreme case, are also given for completeness and comparison.

## 1. Surface UV radiation on Mars and biological implications

The solar UV radiation reaching the Martian surface was computed using a parameterized radiative transfer model (see Appendix) applied to Martian conditions (Córdoba-Jabonero *et al.* 2003). We calculated the instantaneous (over 1 second) UV irradiance for a Martian nominal dust scenario (dust optical depth $\tau_{dust}$ = 0.3) (Lewis *et al.* 1999), corresponding to atmospheric clear-sky conditions on Mars (Schuerger *et al.* 2003), and then instantaneous DNA-weighted dose (also called DNA-weighted irradiance over 1 second) using its action spectrum at local noon (see Appendix).

We note that because the day length ($\Lambda$) and the obliquity ($\gamma$) on Mars are similar to those of Earth ($\Lambda_{Mars}$ / $\Lambda_{Earth}$ = 1.03, $\gamma_{Mars}$ / $\gamma_{Earth}$ = 1.1), the ratio between Martian and terrestrial instantaneous DNA-weighted doses is equivalent to the ratio of daily integrated DNA-weighted doses. Because of the longer Martian year ($T_{Mars}$ / $T_{Earth}$ $\cong$ 1.9), the ratio between seasonally- (or

yearly-) integrated surface DNA-weighted doses on Mars and Earth is enhanced by a factor of < 3. This ratio is small enough to allow us to extrapolate our results from instantaneous to short- or long-term integrated DNA-weighted doses. When comparing terrestrial surface DNA-weighted doses with Martian subsurface DNA-weighted doses, the validity of such approximation is no longer obvious due to the effect of ice refraction. We will address this point in section 3. Thus, the Martian DNA-weighted dose for biological assessments in relation with the terrestrial DNA-weighted dose ($\tau_{dust}$ = 0.1 was chosen for Earth's clear-sky conditions, see Appendix) is seasonally evaluated. The Martian year is expressed through the areocentric longitude ($L_s$) (0° - 360°, where $L_s$ = 0° corresponds to northern spring equinox; in the terrestrial case this corresponds to the value of the Julian day $D_j$ = 80, see Appendix).

First, taking into account a comparative spectral analysis on Mars and Earth, we present the spectral distribution of the UV radiation reaching the polar surfaces on both Mars and Earth. Spectral UV irradiances were calculated for polar latitudes on Mars and Earth (shown in Fig. 1a at 75°N), respectively. Even at these latitudes, where the UV levels should be lower due to the higher solar zenith angles, radiation in the UV-B (280-320 nm) and UV-C ($\lambda$ < 280 nm) ranges still reaches the Martian surface (Fig. 1a). This is in stark contrast with the situation on Earth under normal conditions (Fig. 1a), where due to the ozone layer, the UV-C and part of the UV-B do not reach the terrestrial surface.

In Fig. 1b, the corresponding contour plot for instantaneous DNA-weighted spectral dose is shown for Mars and Earth at 75°N (under normal conditions). The DNA-weighted spectral dose is far more severe for Mars. It is far beyond the terrestrial level not only in magnitude, but also in the spectral range, which is centered mainly between 270 - 290 nm during all seasons, and with spectral components at $\lambda$ < 270 nm (UV-C). The terrestrial DNA-weighted dose is kept within the range of 300-325 nm (UV-B and UV-A) (see Fig. 1b).

The biological damage distribution for Mars is shown in Fig. 2, where a spatial and temporal mapping of instantaneous DNA-weighted dose was calculated by integrating the DNA-weighted spectral irradiance over lambda at each latitude and $L_s$ at local noon. The nearly temporal symmetry of the terrestrial DNA-weighted dose is not observed in the Martian case due to the higher eccentricity ($e$) of this planet ($e_{Mars}$ / $e_{Earth}$ = 5.8). Consequently, the Martian southern hemisphere is exposed to higher solar radiation levels at local noon (which would thus cause greater DNA damage) than the northern hemisphere at the equivalent season time, considering the atmospheric conditions (vertical gas profile and $\tau_{dust}$) to be uniform over the whole planet. During southern summer, Mars reaches perihelion and the southern hemisphere receives a correspondingly higher solar insolation due to its proximity to the Sun, when compared to northern summer which occurs towards aphelion.

The maximal instantaneous DNA-weighted dose on Mars is calculated to be 116 BDU (Biological Dose Unit, see Appendix) at local noon, which is ~ 800 times greater than that under normal terrestrial UV conditions (0.15 BDU, see Appendix) and ~ 200 times higher than the one under terrestrial extreme, ozone-depleted but bearable, UV conditions (0.60 BDU, see Appendix). However, there are regions on Mars with a DNA-weighted dose within the terrestrial "habitable" limit (i.e., maximal terrestrial DNA-weighted dose under normal ozone conditions: 0.15 BDU). These regions relate to the seasonal "night" period, occurring in the autumn and winter seasons for polar latitudes (> 60°N or °S). Furthermore, during this polar "night" period temperatures fall to -128°C (Forget and Pollack 1996) and there is of course no VIS light. For this reason, we are interested in other seasonal times, spring and summer, when solar radiation is present.

## 2. Shielding of UV radiation in polar ice-covered environments

As mentioned previously, from the point of view of liquid water availability and UV and VIS radiation, the polar regions are good candidates for life to survive and grow. During autumn and winter at the Martian surface the DNA-weighted dose is below the highest terrestrial values. But during spring and summer the DNA-weighted dose is higher the range of values found on Earth. However other protection mechanisms are available at the Martian poles such as shielding by the polar ice covers. Snows and ices on Earth do provide habitats for phototrophs (Hoham 1975; Thomas and Duval 1995; Takeuchi 2001), allowing for bacterial and algae metabolism to occur at temperatures below zero (Rivkina *et al.* 2000; Jakosky *et al.* 2003; Price and Sowers 2004). Furthermore ancient microorganisms might be preserved in Martian ices and snows in an analogous way to glacial ices on Earth (Christner *et al.* 2000).

On Mars, a seasonally varying surface layer of $CO_2$ ice exists in the polar regions (Forget 1998; James *et al.* 2000, James and Cantor 2001; Smith et al. 2001) which gradually decreases from winter through summer leaving an exposed perennial $H_2O$ ice layer (James and Cantor 2001; Bibring *et al.* 2004).

The seasonal variation of the $CO_2$ ice cover was obtained from the European Martian Climate Database model (Lewis *et al.* 1999). This study takes into account the seasonal freezing-thawing and related $CO_2$ sublimation/condensation processes (Forget 1998). We then converted the $CO_2$ ice cover (kg m$^{-2}$) to thickness of the $CO_2$ ice layer using the estimated density for the $CO_2$ ice in the Martian caps, 910 kg m$^{-3}$ (Smith *et al.* 2001). An example of the $CO_2$ ice thickness as function of season, along the longitude of 30°E, is shown in Fig. 3. The experimental altimetry results obtained by Mars Orbital Laser Altimeter (MOLA, Mars Global Surveyor mission) (Smith *et al.* 2001) for certain locations in the polar caps are consistent with this mapping.

We next examined whether these ice covers are good candidates, from the point of view of UV radiation, to shelter life. Below the $CO_2$ ice cover, for comparison, we have considered two possible morphologies of the permanent $H_2O$ ices (see Appendix). Fig. 4 shows the absorption coefficients for $CO_2$ ice (Hansen 1997), $H_2O$ ice and $H_2O$ snow (Perovich 1993) in the UV and VIS ranges. The absorption of $H_2O$ ice and $H_2O$ snow is comparable, whereas the scattering is higher for the snow (Fig. 4 shows only the absorption coefficients). They both have stronger UV-VIS absorption than $CO_2$ ice throughout the whole range. Radiative transfer through ice covers was performed by using a parameterized model based on the UV propagation through a layered medium with different optical properties in each layer (see Appendix).

In Fig. 5a we show the depth of $CO_2$ ice needed to reduce the DNA-weighted dose to values similar to those found on Earth, in spring and at northern polar latitudes ($L_s$ = 30°). The required depth (several meters, see Fig. 5a) is far greater than the reported existing $CO_2$ ice depth, $\sim 1$ m (Forget 1998; Smith *et al.* 2001). Thus the seasonally varying $CO_2$ ice cover is insufficient to reduce the biologically harmful UV radiation to terrestrial levels, and therefore the presence of $H_2O$ ice would be needed to reduce UV levels to terrestrial values. We have done similar calculations for the $H_2O$ ice and snow at the same latitudes (see Fig. 5b and 5c, respectively), and we conclude that depths of the order of one meter of $H_2O$ ice or a few centimeters of $H_2O$ snow are sufficient to obtain terrestrial DNA-weighted doses.

## 3. Martian *Radiative Habitable Zone* (RHZ)

In order to determine the *Radiative Habitable Zone* (RHZ) in Mars, we have studied the radiative conditions under polar ice-covered surfaces based on Earth-like radiative "habitability" criteria (see Appendix). In particular if we consider photosynthetic microorganisms, in addition to the DNA-weighted dose, we need to consider sufficient transmittance of *Photosynthetically Active Radiation* (PAR) through the ice cover. PAR is the optically visible (VIS) part of the solar spectrum reaching the surface/subsurface, given in moles of photons $s^{-1}$ $m^{-2}$. The VIS radiation, which may be used as energy source by photosynthetic organisms, cannot be lower than a minimal value ($PAR_{min}$) necessary to sustain biological processes. A good review of low PAR levels is presented in Raven *et al.* (2000), where a value of $PAR_{min} = 100$ nmoles $s^{-1}$ $m^{-2}$ was calculated. However, we were interested in the lowest allowed PAR limit. Therefore, a value of $PAR_{min} = 10$ nmoles $s^{-1}$ $m^{-2}$ was chosen for the minimal PAR condition, which corresponds to the observed value obtained by Litter *et* *al.* (1986) for red macro-algae photolithotrophically growing at 274 m ocean water depth.

Thus, as it has been said before, for photosynthetic microorganisms a region is "radiatively habitable" if it fulfills two radiative conditions: 1) the DNA-weighted dose ($D_{Mars}$) must be equal to or lower than the terrestrial DNA-weighted dose ($D_{Earth}$) under normal conditions; and 2) PAR must be higher than $PAR_{min}$. Those regions on Mars satisfying both these two conditions are classed here as *Radiative Habitable Zone* (RHZ). Thus the RHZ is comprised between two boundaries.

The upper radiative limit is defined by the minimal depth ($z_{min}$) at which the Martian DNA-weighted dose ($D_{Mars}(z)$) is lower than the terrestrial dose under normal conditions ($D_{Earth}$). In order to calculate the $D_{Mars}(z)$ as function of ice depth, we evaluate the radiation propagation from the surface to a given ice depth ($z$). The "total" ice depth includes a seasonally dependent $CO_2$ ice cover (see Fig. 3) lying on top of a perennial and sufficiently deep $H_2O$ ice or snow layer, which is assumed to exist only at latitudes greater than 75°N or °S. We obtain thus a "mapping" of the upper limit data ($z_{min}$) as a function of latitude and $L_s$ (Martian year). Fig.6 and 7 (left column) show $z_{min}$ for the north and south poles, respectively. In all the cases a nominal dust scenario ($\tau_{dust}$ = 0.3, clear-sky conditions) is considered, but similar results were obtained for a slightly elevated dust scenario ($\tau_{dust}$ = 0.6, moderate-dusty-sky conditions) (data not shown). Certain characteristic depths ($z_1$ to $z_{10}$), whose meaning will be next defined, are summarized in Table 1. For the case of $H_2O$ ice, at "total" depths $z$ > $z_1$ the environment is protected from UV radiation *throughout the whole*

*Martian year* (see Fig. 6a and 7a). This corresponds to depths, within the permanent $H_2O$ ice layer, greater than $z_2$. The same depths for the case of $H_2O$ snow are $z_3$ and $z_4$, respectively (see Fig. 6c and 7c).

The lower radiative limit is defined by the maximal depth ($z_{max}$) where the Martian PAR is higher than the $PAR_{min}$, required for known photosynthetic organisms. Similar calculations have been performed for the transmission of the visible part of the spectrum, and a "mapping" of the $z_{max}$ data is obtained under a nominal dust scenario ($\tau_{dust}$ = 0.3). For the $H_2O$ ice case, the PAR level is sufficiently high *all along the spring and summer time* (during the "polar night", autumn and winter, no radiation gets to the surface either) up to total depths of $z_5$, whereas lower depths within the ice receive acceptable levels of PAR during fewer days. Beyond $z_6$ total depth the PAR level is *never sufficient* for photosynthesis to take place. For the $H_2O$ snow case, due to its stronger scattering properties, this last defined depth is reduced to a few centimeters ($z_7$) within the permanent $H_2O$ snow cover. In Fig. 6 and 7 (right column) we show a mapping of the RHZ thickness ($z_{max}$ - $z_{min}$) for the case of $H_2O$ ice (Fig. 6b and 7b) and $H_2O$ snow (Fig. 6d and 7d). This thickness varies during the Martian northern and southern spring and summer between $z_8$ and $z_9$ for $H_2O$ ice-like layers, and a few centimeters ($z_{10}$) for snow-like morphology, depending strongly on the optical properties of the permanent $H_2O$ ices.

For completeness we evaluated the ratio of the Martian subsurface to terrestrial surface yearly integrated DNA-weighted doses: taking into account the longer Martian year ($T_{Mars}$ / $T_{Earth}$ $\cong$ 1.9), and the corresponding ice refraction indexes, the ratio of yearly integrated DNA-weighted doses is only in the range of 0.8 to 1.1 times greater than the ratio of instantaneous DNA-weighted doses. Therefore, we have verified that determining the RHZ by using the instantaneous DNA-weighted dose is a good approximation to the RHZ computed with the integrated DNA-weighted dose in order to define Earth-like "habitability". This is especially true at high latitudes where Earth-like life forms living in the "instantaneous RHZ" would not receive in continuous more than ~ 1.7 times the DNA-weighted dose received in continuous by a terrestrial organism.

Previous theoretical studies on $H_2O$ snow-pack covers in the Martian north pole (80°N, local noon, summer solstice) reported an upper radiative limit $z_{min}$ of 18 cm (for UV protection) and lower radiative limit $z_{max}$ of 30 cm (because of PAR requirements) (Cockell and Raven 2004). These results were based on field measurements of UV radiation transmission through Antarctic natural snows to characterize its optical properties. The difference in the RHZ is due to the fact that the snow scattering properties used in this fore mentioned study correspond to an intermediate scenario, between the $H_2O$ ice-like and snow-like cases considered here. Unfortunately the detailed morphological characteristics (snow versus ice) and composition (presence of dust, mixture of $H_2O$ / $CO_2$ ices, …) of the Martian polar caps is still unknown, but the three scenarios considered here may serve to estimate the dependence of the RHZ with the optical properties of the perennial ice layer.

## 4. Discussion and Conclusions

Searching for potentially habitable UV-protected environments, we have considered the Martian polar ice caps. In particular, the northern polar region, which is a more favorable environment for life in many aspects, is studied in detail and analogous results are reported for the southern polar region. Beneath the seasonally varying $CO_2$ ice layer, there is a perennial $H_2O$ ice layer, which is left uncovered in summer when the $CO_2$ sublimates. This $H_2O$ ice layer is a good candidate to shelter life due to its permanence and the UV absorption properties of the $H_2O$ ice. It was found that the deposited $CO_2$ ice layer is not sufficient to screen the UV radiation by itself. Nevertheless, at $\sim 1$ m depth in the perennial $H_2O$ ice layer the DNA-weighted dose is reduced to levels similar to those found on Earth. At up to $\sim 4$ m depth VIS radiation reaches these UV-protected regions with sufficient intensity for photosynthesis to take place. Moreover, the RHZ thickness (difference between the lower and upper radiative limits) varies between 1.5 m and 2.4 m during the Martian year. Similar calculations with $H_2O$ snow reduce the RHZ to a few centimeters (2 - 4 cm). Thus the RHZ depends strongly on the morphological properties of the perennial $H_2O$ ice layer. Absorption and refraction through the ice make the relationship between the instantaneous and integrated DNA-weighted dose much more complex than at the surface. We have verified that the RHZ determined using the instantaneous DNA-weighted dose is a good approximation to the RHZ computed with the integrated DNA-weighted dose in order to define Earth-like "radiative habitability". Similar results were obtained for both poles, although with slightly deeper radiative limit layers in the southern case.

One of the main unknowns about the characteristics of the Martian polar environment is the morphology (and related optical properties) of the $CO_2$ and $H_2O$ ice polar caps. The snow and ice morphologies define two limiting scenarios of the polar caps optical properties. In our model for the $H_2O$ polar caps, we have chosen the so-called "interior of white ice" for $H_2O$ ice and "wet snow" for $H_2O$ snow, which have been successfully used to characterize the Antarctic snow-ice covers (Perovich 1993; Cockell and Córdoba-Jabonero 2004). The absorption is comparable in both cases, while the scattering is higher for the snow case (see Appendix). However in the case of $CO_2$, since we are not aware of any radiative transmission study on $CO_2$ snows, we have only considered here the $CO_2$ ice morphology for which the optical properties are known (Hansen 1997). If the Martian $CO_2$ ice caps were more similar to snow then the UV flux would be strongly attenuated due to the stronger scattering properties of snow as compared to ice (as in the case of $H_2O$). Nevertheless, because of the rapid sublimation of $CO_2$ during spring, only the perennial $H_2O$ ice layer may provide an effective shield along the whole year.

There are still many other unknowns about this environment such as the thickness of the perennial $H_2O$ ice and its dynamics, local temperatures, the concentration of dust and salts in the ice and its influence on the optical and thermodynamical properties (in particular, conditions required for liquid water to be formed at polar ice layers below the surface), the permafrost characteristics and the seasonal atmospheric local dynamics in this region. Further refinements to the radiative transfer model may vary the radiative habitable zone (RHZ).

The scientific community has shown great interest on the Martian polar regions for future science explorations (Clifford *et al.* 2000). This study allowed us to define eligible regions and ice depths to be explored by future missions searching for radiative habitable environments. We propose the subsurface of the perennial Martian ice caps as a possible target for future Martian lander missions searching for extant or extinct living forms. Furthermore we should keep in mind that these regions may also be susceptible of contamination by Earth-like life forms, and special attention should be paid for planetary protection.

**APPENDIX: Characterization of the solar radiation levels on Earth and Mars**

*A. Radiative Transfer Modeling*

The mean distances of Mars and Earth to the Sun are 1.524 and 1 AU respectively, and therefore the mean solar flux at the top of the Martian atmosphere is only 43% that of Earth. The seasonal variations of the Martian solar flux are taken into account by correcting the flux at 1 AU (Nicolet 1989) with the varying Martian distance to the Sun as a function of areocentric longitude ($L_s$).

To characterize the UV biological impact we used the biologically weighted irradiance, also called the instantaneous (over 1 second) biologically weighted dose, $D(z)$, which was calculated by convolution of the spectral irradiance or spectral flux $F(\lambda, z)$ reaching the surface ($z = 0$) or a given depth $z$ in the subsurface, with the corresponding Biological Action Spectrum, $B(\lambda)$, integrating over $\lambda$, as follows:

$$D(z) = \Sigma_\lambda \, D(\lambda, \, z) \, \Delta\lambda = \Sigma_\lambda \, F(\lambda, \, z) \times B(\lambda) \times \Delta\lambda \qquad\qquad \text{(A1)}$$

where $D(\lambda, \, z)$ is the spectral biologically weighted irradiance (or instantaneous over 1 second spectral biologically weighted dose). The symbol $\Sigma_\lambda$ represents the spectral sum in the UV range from 200 to 400 nm (UV-C + UV-B + UV-A), and $\Delta\lambda = 1$ nm is the wavelength increment.

$B(\lambda)$ characterizes the sensitivity or biological response of an organism to UV radiation. We chose DNA as a representative biological target, since it has a well studied $B(\lambda)$ (Setlow and Doyle 1954; Setlow 1974; Green and Miller 1975; Lindberg and Horneck 1991; Horneck 1993) and is widely used for biological implication assessments in both terrestrial (e.g., Madronich 1993; Córdoba-Jabonero 2003) and Martian studies (e.g., Cockell *et al.* 2000; Córdoba-Jabonero *et al.* 2003; Patel *et al.* 2004). Although the biologically weighted irradiance (or instantaneous biologically weighted dose) is commonly given in W m$^{-2}$ (second is usually omitted), so it has not the "dimension" of an energy flux. Indeed, the biological response $B(\lambda)$ is not a "strict" weight function normalized over a wavelength range, but is commonly given relatively to its value at 300 nm. As a consequence, one can find, for a given wavelength range, instantaneous biologically weighted doses $D(z)$ in W m$^{-2}$ exceeding the spectrally integrated energy flux $F(z)$ itself. To avoid such inconsistency, we give $D(z)$ (integrated over 1 second) in BDU (Biological Dose Unit) but the reader should keep in mind that these values can be quantitatively compared with instantaneous biological doses usually given in W m$^{-2}$ in the literature.

Action spectra below 280 nm are not generally obtained since UV radiation at wavelengths shorter than 280 nm is completely removed by the Earth's atmosphere. We used a generalized DNA action spectrum $B(\lambda)$, which was obtained by convolving the DNA absorbance from 200 to 280 nm (Lindberg and Horneck 1991) with a standard DNA action spectrum (Setlow 1974) at wavelengths $\lambda$ >280 nm. Thus, DNA damage peaks in the UV-C and part of the UV-B ranges, as shown in Fig. 8.

The solar UV and VIS irradiance $F(\lambda, z)$ reaching the surface ($z = 0$) of Mars and Earth was computed using a parameterized radiative transfer model, where the direct component of radiation is described by the Lambert-Beer law and the diffuse component is simulated on the basis of the *two-stream* approximation (Iqbal 1983). We consider first the radiation propagating forward through the atmosphere, which is modeled here as a plane parallel layered homogeneous medium with a thickness of 200 km. Each atmospheric layer is defined with different absorbing and scattering properties dependent upon the vertical distribution of the atmospheric composition. This is introduced into the model by the mass factor of each component, depending on the solar zenith angle (SZA), and corrected for the atmospheric refraction index and the curvature of the planet. This correction is very important for high SZAs. The SZA is calculated using the geometric (latitude and longitude of location, solar time, etc.). and orbital (eccentricity, obliquity, orbital time, etc.) parameters for each planet. Both the direct and diffuse radiation components are evaluated on a horizontal plane at the surface.

In the case of the Earth, as used in (Córdoba-Jabonero 1999; Córdoba-Jabonero *et al.* 2004), $O_2$ and $O_3$ are the main absorbers in the UV and VIS range, but $O_3$ is the only molecule inducing significant spatial and temporal variations in the surface UV irradiation. The content of $O_3$ is defined by the Total Ozone Column (TOC). Rayleigh scattering is taken for the air and the aerosol scattering contribution is determined by the aerosol optical thickness for the dust $\tau_{dust}$. This parameter is wavelength dependent as defined by the Angstrom turbidity equation (Angstrom 1929), thus the values of $\tau_{dust}$ in this work will be referred to $\lambda$ = 550 nm. This model has been successfully validated with UV measurements at the Earth's surface under different atmospheric conditions and at different geographical locations (Córdoba-Jabonero *et al.* 2004). The terrestrial year length ($T_{Earth}$), is expressed by the Julian day ($D_j$), being $D_j$ = 1 the 1$^{st}$ of January, ... and $D_j$ = 365 the 31$^{st}$ of December. In the case of Mars, the solar absorption is due to the $CO_2$, $O_2$ and $O_3$. The vertical profiles and mixing ratios are the same as used in (Córdoba-Jabonero *et al.* 2003, see there Fig. 1). Martian $\tau_{dust}$ (at $\lambda$ = 550 nm) is also obtained by the Angstrom turbidity equation.

Next, for the studies within the Martian ice caps, $F(\lambda, z)$ on the subsurface was calculated by considering the radiative transfer through a layered medium in a similar way to previous studies performed for aquatic media (Córdoba-Jabonero *et al.* 2004). Here the absorbing and scattering properties of $CO_2$ ice and $H_2O$ snow/ice define the radiative characteristic of this layered medium. $CO_2$

absorption coefficients are taken from Hansen (1997). We have used the $H_2O$ ice extinction (absorption and scattering) coefficients, also those for $H_2O$ snow, given in Perovich (1993), which are based on observations in Arctic and Antarctica, laboratory measurements and interpolations. In our model for the Martian polar caps, we have chosen the so-called "interior of white ice" (5-type) for $H_2O$ ice and "wet snow" (2-type) for $H_2O$ snow, which have been successfully used to characterize the Antarctic snow-ice covers (Cockell and Córdoba-Jabonero 2004). The absorption is comparable in both cases, while the scattering is quite higher for the snow case (Fig. 4 shows only the absorption coefficients). Besides, in the modeling calculations we have taken into account the refraction index variation through the atmosphere, $CO_2$ ice and $H_2O$ ice layers.

*B. Terrestrial instantaneous DNA-weighted dose (or DNA-weighted irradiance) as an environmental radiative parameter for habitability conditions*

As described above, the terrestrial spectral irradiance is obtained using the following variable parameters: Total Ozone Column (TOC), aerosol optical thickness for dust $\tau_{dust}$, and solar zenith angle (SZA). In this work we are only considering limitations to habitability related to the solar radiation exposure, and thus we regard the terrestrial surface instantaneous DNA-weighted dose ($D_{Earth}$) as the critical parameter to define environmental radiative habitability. On present-day Earth, the change of TOC is the main factor leading to a variation of global UV radiation levels, and consequently variability of DNA-weighted doses. The enhancement of solar UV radiation and DNA-weighted dose due to ozone depletion can reach high levels. This occurs, for instance, in Antarctica during the period of "ozone hole" formation each austral spring. We will therefore distinguish between two possible values of DNA-weighted dose, which occur on Earth under different UV environmental conditions: normal conditions with a standard TOC (350 Dobson Units, DU) and extreme, but still bearable, UV conditions (e.g., Antarctic-like conditions with a 50% ozone depletion: 175 DU).

We evaluated the biological damage distribution under each ozone condition mapping spatially (in latitude) and temporally (during the orbital year) the DNA-weighted dose for Earth. We considered two hypothetical scenarios with homogeneous ozone depletion in time and latitude. These DNA-weighted dose values are calculated for the worst radiation conditions: a terrestrial low dust scenario ($\tau_{dust}$ = 0.1, Earth's clear-sky conditions) (Holben *et al.* 1998) and at local noon. Therefore, the critical habitability maximal values for $D_{Earth}$ are: 0.15 BDU for normal conditions, and 0.60 BDU for extreme conditions (4 times greater). It is worth noting here that VIS radiation is not affected by ozone depletion. The VIS radiation at noon ranges throughout the year between 300 W m$^{-2}$ and 520 W m$^{-2}$ at equatorial latitudes and between 0.5 W m$^{-2}$ and 280 W m$^{-2}$ towards the polar regions (e.g., 75°N, data not shown), excluding the "polar night" period.

**Acknowledgments**

CCJ gratefully acknowledges the support by the Consejería de Educación de la Comunidad de Madrid (CAM, Spain) for providing a postdoctoral research contract. The work of MPZ and FS is supported by the Instituto Nacional de Técnica Aeroespacial. MRP acknowledges funding from PPARC as part of post-doctoral support. We used the European Martian Climate Database for the $CO_2$ ice cover data.

**REFERENCES**

Angstrom, A., 1929. On the atmospheric transmission of sun radiation and on dust in the air. Geographys. Annal. 2, 156-166.

Baumstark-Khan, C., Facius, R., 2002. Life under condition of ionizing radiation. In: Horneck, G., Baumstark-Khan, C. (Eds.), Astrobiology: The Quest for the Conditions of Life, Springer-Verlag, Berlin, Heidelberg, New York, pp. 261-284.

Bibring, J.-P., Langevin, Y., Poulet, F., Gendrin, A., Gondet, B., Berthe, M., Soufflot, A., Drossart, P., Combes, M., Bellucci, G., Moroz, V.I., Mangold, N., Schmitt, B., and The OMEGA Team, 2004. Perennial water ice identified in the south polar cap of Mars. Nature 428, 627-630.

Carpenter, E.J., Senjie, L., Capone, D.G., 2000. Bacterial activity in South Pole snow. Applied and environmental microbiology 6 (10), 4514-4517.

Christner, B.C., Mosley-Thompson, E., Thompson, L.G., Zagorodnov, V., Sandman, K., Reeve, J.N., 2000. Recovery and identification of viable bacteria immured in glacial ice. Icarus 44 (2), 479-485.

Clifford, S.M., 53 colleagues, 2000. The state and future of Mars polar science and exploration. Icarus 144, 210–242.

Cockell, C.S., Córdoba-Jabonero, C., 2004. Coupling of climate change and biotic UV exposure through changing snow-ice covers in terrestrial habitats. Photochem. Photobiol. 79 (1), 26-31.

Cockell, C.S., Raven, J.A., 2004. Zones of photosynthetic potential on Mars and the early Earth. Icarus 169, 300-310.

Cockell, C.S., Catling, D.C., Davis, W.L., Snook, K., Kepner, R.L., Lee, P., McKay, C.P., 2000. The ultraviolet environment of Mars: Biological implications. Past, present, and future. Icarus 146, 343-359.

Cockell, C.S., Rettberg, P., Horneck, G., Patel, M.R., Lammer, H., Córdoba-Jabonero, C., 2002. Ultraviolet protection in microhabitats – lessons from the terrestrial poles applied to Mars. In: Proc. Second European Workshop on Exo/Astrobiology, ESA SP-518, pp. 215-218.

Córdoba-Jabonero, C., 1999. La Radiación ultravioleta solar: Métodos de medida y su relación con la capa de ozono. Ph.D. thesis. Universidad Autónoma de Madrid, Spain.

Córdoba-Jabonero, C., 2003. Behaviour of biological ecosystems under extreme environmental solar UV-B radiation. In: Celnikier, L.M., Tran Thanh Van, J. (Eds.), Frontiers of Life, Proceedings of the XXII[th] Rencontres of Blois., The Giòi Publishers, Vietnam, pp.137-139.

Córdoba-Jabonero, C., Lara, L.M., Mancho, A.M., Márquez, A., Rodrigo R., 2003. Solar ultraviolet transfer in Martian atmosphere: biological and geological implications. Planet. Sp. Sci. 51, 399-410.

Córdoba-Jabonero, C., Mancho, A.M., González-Kessler, C., Martín-Soler, J., Jaque, F., 2004. Iron shielding of ultraviolet radiation in a natural extreme acidic aquatic environment: the Tinto river (Spain). J. Geophys. Res. (submitted).

Costard, F., Forget, F., Mangold, N., Peulvast, J.P., 2002. Formation of recent Martian debris flows by melting of near-surface ground. Science 295, 110-113.

Forget, F., Pollack, J.B., 1996. Thermal infrared observations of the condensing Martian polar caps: $CO_2$ ice temperatures and radiative budget. J. Geophys. Res. 101 (E7), 16,865-16,879.

Forget, F., 1998. Mars CO2 ice polar caps. In: Schmitt, B., De Bergh, C., Festou, M. (Eds.), Solar system ices, Kluwer Academic, pp. 477-507

Gilichinsky, D.A., 2002. Permafrost model of extraterrestrial habitat. In: Horneck, G., Baumstark-Khan, C. (Eds.), Astrobiology: The Quest for the Conditions of Life, Springer-Verlag, Berlin, Heidelberg, New York, pp. 271-295.

Green, A.E.S., Miller, J.H., 1975. Measures of biologically effective radiation in the 280–340 nm region. CIAP Monogr. 5 (1), 2.60-70.

Haberle, R.M., McKay, C.P., Schaeffer, J., Cabrol, N.A., Grin, E.A., Zent, A.P., Quinn, R., 2001. On the possibility of liquid water on present-day Mars. J. Geophys. Res. 106 (E10), 23317-23326.

Hansen, G.B., 1997. Spectral absorption of solid $CO_2$ from the ultraviolet to the far-infrared. Adv. Space Res. 20 (8), 1613-1616.

Hecht, M.H., 2002. Metastability of liquid water on Mars. Icarus 156 (2), 373-386.

Hoham, R.W., 1975. Optimum temperatures and temperature ranges for growth of snow algae. Arctic and Alpine Res 7, 13-24.

Hoham, R.W., Mullet, J.E., Roemer, S.C., 1983. The life history and ecology of the snow alga *Chloromonas polyptera* comb. nov. (Chlorophyta, Volvocales). Canadian J. Botany 61, 2416-2429.

Holben, B.N., Eck, T.F., Slutsker, I., Tanré, D., Buis, J.P., Setzer, A., Vermote, E.F., Reagan, J.A., Kaufman, Y.J., Nakajima, T., Lavenue, F., Jankowiak, I., Smirnov, A., 1998. AERONET - A federated instrument network and data archive for aerosol characterization. Remote Sens. Environ. 66, 1-16.

Horneck, G., 1993. Responses of *Bacillus subtilis* spores to space environment: Results from experiments in space. Origins Life Evol. Biosphere 23, 37-52.

Iqbal, M., 1983. An Introduction to Solar Radiation. Academic Press, Ontario, Canada.

Jakosky, B.M., Nealson, K.H., Bakermans, C., Ley, R.H., Mellon, M.T., 2003. Subfreezing activity of microorganisms and the potential habitability of Mars' polar regions. Astrobiology 3 (2), 343-350.

James, P.B., Cantor, B.A., Malin, M.C., Edgett, K., Carr, M.H., Danielson, G.E., Ingersoll, A.P., Davies, M.E., Hartman, W.K., McEwen, A.S., Soderblom, L.A., Thomas, P.C., Veverka, J., 2000. The 1997 spring regression of the Martian south polar caps: Mars Orbiter Camera observations. Icarus 144, 410-418.

James, P.B., Cantor, B.A., 2001. Martian north polar cap recession: 2000 Mars Orbiter Camera observations. Icarus 154, 131-144.

Kuhn, W.R., Atreya, S.K., 1979. Solar radiation incident on the Martian surface. J. Mol. Evol. 14, 57–64.

Lewis, S.R., Collins, M., Read, P.L., Forget, F., Hourdin, F., Fournier, R., Hourdin, C., Talagrand, O., Huot, J.-P., 1999. A climate database for Mars. J. Geophys. Res. - Planets 104 (E10), 24,177-24,194 (http://www-mars.lmd.jussieu.fr).

Lindberg, C., Horneck, G., 1991. Action spectra for survival and spore product formation of *Bacillus subtilis* irradiated with short wavelength (200–300 nm) UV at atmospheric pressure and *in vacuo*. J. Photochem. Photobiol. 11, 69-80.

Littler, M.M., Littler, D.S., Blair, S.M., Norris, J.N., 1986. Deep-water plant communities from an uncharted seamount off San Salvador Island, Bahamas: distribution, abundance and primary production. Deep-see Res. 33, 881-892.

Lobitz, B., Wood, B.L., Averner, M.M., McKay, C.P., 2001. Use of spacecraft data to derive regions on Mars where liquid water would be stable. Proceed. Nat. Acad. Sci. 98, 2132-2137.

Madronich, S., 1993. The atmosphere and UV-B radiation at ground level. In: Young, A.R., Bjorn, L.O., Moan, J., Nultsch, W. (Eds.), Environmental UV Photobiology, Plenum Press, New York, pp. 1-39.

Matson, D.L., Brown, R.H., 1989. Solid-state greenhouses and their implications for icy satellites. Icarus 77, 67-81.

Nicolet, M., 1989. Solar spectral irradiances and their diversity between 120 and 900 nm. Planet. Space Sci. 37, 1249-1289.

Patel, M.R., Zarnecki, J.C., Catling, D.C., 2002. Ultraviolet radiation on the surface of Mars and the Beagle 2 UV Sensor. Planet. Sp. Sci. 50 (9), 915-927.

Patel, M.R., Christou, A.A., Cockell, C.S., Ringrose, T.J., Zarnecki, J.C., 2004. The UV environment of the Beagle 2 landing site: Detailed investigations and detection of atmospheric state. Icarus 168 (1), 93-115.

Perovich, D.K., 1993. A theoretical model of ultraviolet light transmission through Antarctic sea ice. J. Geophys. Res. 98 (C12), 22579-22587.

Price, P.B., Sowers, T., 2004. Temperature dependence of metabolic rates for microbial growth, maintenance, and survival. Proceed. Nat. Acad. Sci. 101 (13), 4631-4636.

Raven, J.A., Kübler, J.E., Beardall, J., 2000. Put out the light, and then put out the light. J. Mar. Biol. Ass. UK 80, 1-25.

Rivkina, E.M., Friedmann, E.I., McKay, C.P., Gilichinsky, D.A., 2000. Metabolic activity of permafrost bacteria below the freezing point. Appl. Environ. Microbio., 66, 3230-3233.

Sagan, C., Pollack, J.B., 1974. Differential transmission of sunlight on Mars: Biological implications. Icarus 21 (4), 490-495.

Schuerger, A.C., Mancinelli, R.L., Kern, R.G., Rothschild, L.J., McKay, C.P., 2003. Survival of endospores of *Bacillus subtilis* on spacecraft surfaces under simulated martian environments: implications for the forward contamination of Mars. Icarus 165, 253-276.

Setlow, R.B., 1974. The wavelengths in sunlight effective in producing skin cancer: a theoretical analysis. Proc. Natl. Acad. Sci., USA 7, 3363-3366.

Setlow, R., Doyle, B., 1954. The action of radiation on dry Deoxyribonucleic acid. Biochim. Biophys. Acta 15, 117-125.

Smith, D.E., Zuber, M.T., Neumann G.A., 2001. Seasonal variations of snow depth on Mars. Science 294, 2141-2146.

Takeuchi, N., 2001 The altitudinal distribution of snow algae on an Alaskan glacier (Gulkana glacier in the Alaska Range). Hydrol. Process. 15, 3447-3459.

Thomas, W.H., Duval, B., 1995. Snow algae: snow albedo changes, algal-bacterial interrelationships, and ultraviolet radiation effects. Arctic Alpine Res., 27, 389-399.

Warren, S.G. 1984. Optical constants of ice from the ultraviolet to the microwave. Appl. Opt. 25, 1206-1225.

**Table 1.** Characteristic ice depths (in m), defined in the text, which determine the Radiative Habitable Zone (RHZ) of the Martian polar regions.

| Polar region | $z_1$ | $z_2$ | $z_3$ | $z_4$ | $z_5$ | $z_6$ | $z_7$ | $z_8$ | $z_9$ | $z_{10}$ |
|---|---|---|---|---|---|---|---|---|---|---|
| °N | 1.4 | 1.0 | 0.9 | ~0.04 | 1.8 | 3.6 | ~0.06 | 1.5 | 2.3 | 0.02-0.04 |
| °S | 1.8 | 1.1 | 1.1 | ~0.04 | 1.8 | 4.0 | ~0.07 | 1.5 | 2.4 | 0.02-0.04 |

**FIGURE CAPTIONS**

**Fig. 1.** Solar UV irradiance (W m$^{-2}$ nm$^{-1}$) (a), and instantaneous DNA-weighted spectral dose (BDU nm$^{-1}$) (b), at local noon, for a polar latitude of 75°N, on Mars and Earth under normal ozone conditions (350 DU). Seasons for the northern hemisphere are indicated.

**Fig. 2.** Instantaneous DNA-weighted dose (BDU) mapping on Mars at local noon and for a nominal dust scenario ($\tau_{dust}$ = 0.3, clear-sky conditions). Seasons for the northern hemisphere are indicated.

**Fig. 3.** Mapping of the $CO_2$ ice thickness (in cm) along the 30°E meridian as a function of season ($L_s$), converted from the $CO_2$ ice cover (kg m$^{-2}$) given by the European Martian Climate Database (Lewis *et al.* 1999). Dashed lines indicate the limits of the perennial and sufficiently deep $H_2O$ ice cover assumed in our model.

**Fig. 4.** Absorption coefficients for $CO_2$ ice (Hansen 1997), $H_2O$ ice and snow (Perovich 1993).

**Fig. 5.** Variation of the Martian DNA-weighted dose ($D_{Mars}$) relative to terrestrial values ($D_{Earth}$) at local noon, for $L_s$ = 30° (northern spring), as a function of the $CO_2$ ice (a), $H_2O$ ice (b) and $H_2O$ snow (c) depths, for different northern latitudes (from 60°N through 90°N in intervals of 5°). For comparison the terrestrial levels of instantaneous (over 1 second) DNA-weighted dose under normal (0.15 BDU, vertical thicker dashed arrow) and extreme, depleted ozone, conditions (0.60 BDU, vertical thinner dashed arrow) are marked.

**Fig. 6.** Mapping of the upper and lower ice depth limits (in cm) of the Martian Radiative Habitable Zone (RHZ) for northern polar regions. The upper radiative limit $z_{min}$ at which $D_{Mars} \leq D_{Earth}$ (left column), and thickness ($z_{max} - z_{min}$ ) (right column), where the lower radiative limit $z_{max}$ is such that $PAR_{Mars} \geq PAR_{min}$, are shown for two polar cap scenarios: ($CO_2$ ice + $H_2O$ ice) cover (a and b) and ($CO_2$ ice + $H_2O$ snow) cover (c and d). Seasons for the northern hemisphere are indicated.

**Fig. 7.** Mapping of the upper and lower ice depth limits (in cm) of the Martian Radiative Habitable Zone (RHZ) for southern polar regions. The upper radiative limit $z_{min}$ at which $D_{Mars} \leq D_{Earth}$ (left column), and thickness ($z_{max} - z_{min}$ ) (right column), where the lower radiative limit $z_{max}$ is such that $PAR_{Mars} \geq PAR_{min}$, are shown for two polar cap scenarios: ($CO_2$ ice + $H_2O$ ice) cover (a and b) and ($CO_2$ ice + $H_2O$ snow) cover (c and d). Seasons for the southern hemisphere are indicated.

**Fig. 8.** Generalized DNA action spectrum (normalized at 300 nm) used in this work (Setlow 1974; Lindberg and Horneck 1991).

**Fig. 1.** Solar UV irradiance (W m$^{-2}$ nm$^{-1}$) (a), and instantaneous DNA-weighted spectral dose (BDU nm$^{-1}$) (b), at local noon, for a polar latitude of 75°N, on Mars and Earth under normal ozone conditions (350 DU). Seasons for the northern hemisphere are indicated.

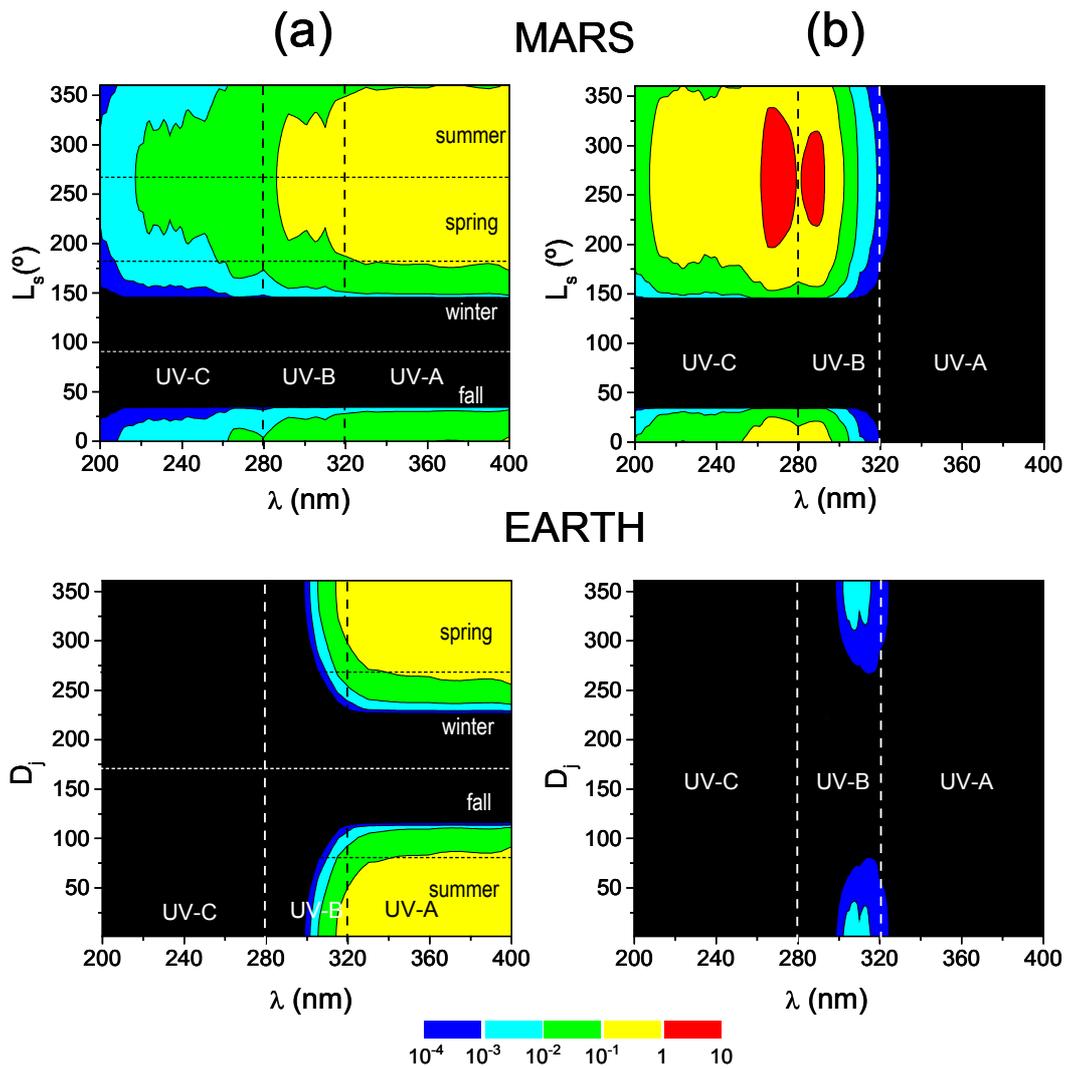

**Fig. 2.** Instantaneous DNA-weighted dose (BDU) mapping on Mars at local noon and for a nominal dust scenario ($\tau_{dust}$ = 0.3, clear-sky conditions). Seasons for the northern hemisphere are indicated.

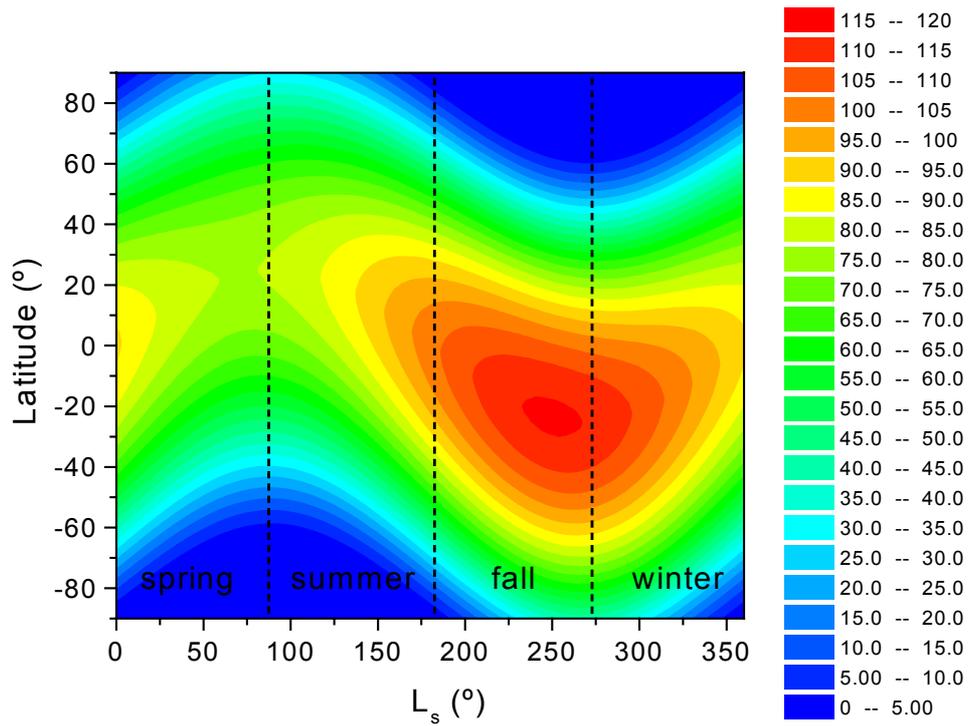

**Fig. 3.** Mapping of the $CO_2$ ice thickness (in cm) along the 30°E meridian as a function of season ($L_s$), converted from the $CO_2$ ice cover (kg m$^{-2}$) given by the European Martian Climate Database (Lewis *et al.* 1999). Dashed lines indicate the limits of the perennial and sufficiently deep $H_2O$ ice cover assumed in our model.

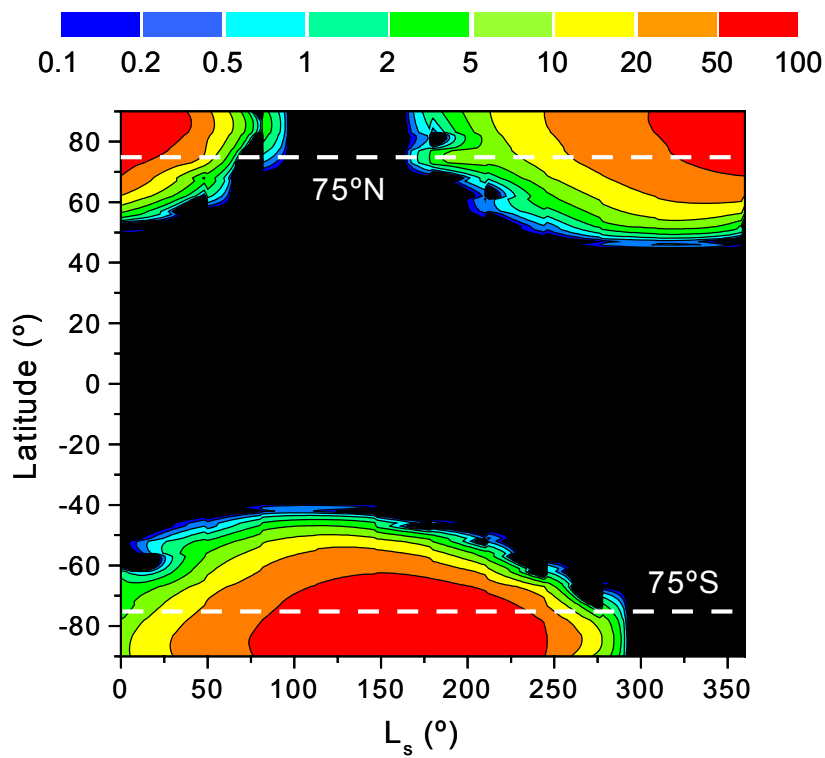

**Fig. 4.** Absorption coefficients for $CO_2$ ice (Hansen 1997), $H_2O$ ice and snow (Perovich 1993).

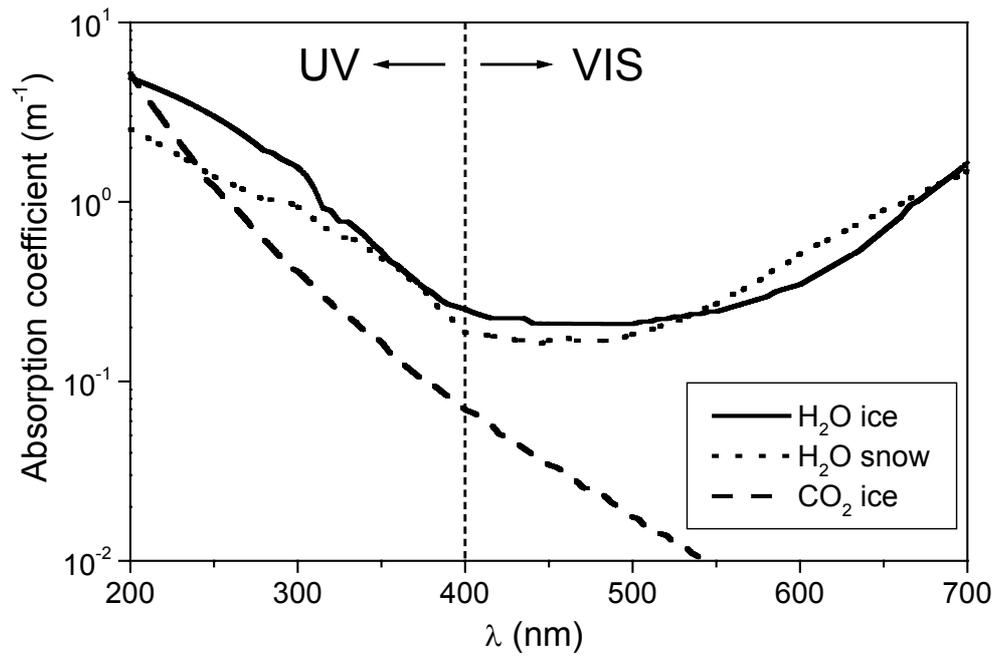

**Fig. 5.** Variation of the Martian DNA-weighted dose ($D_{Mars}$) relative to terrestrial values ($D_{Earth}$) at local noon, for $L_s = 30°$ (northern spring), as a function of the $CO_2$ ice (a), $H_2O$ ice (b) and $H_2O$ snow (c) depths, for different northern latitudes (from 60°N through 90°N in intervals of 5°). For comparison the terrestrial levels of instantaneous (over 1 second) DNA-weighted dose under normal (0.15 BDU, vertical thicker dashed arrow) and extreme, depleted ozone, conditions (0.60 BDU, vertical thinner dashed arrow) are marked.

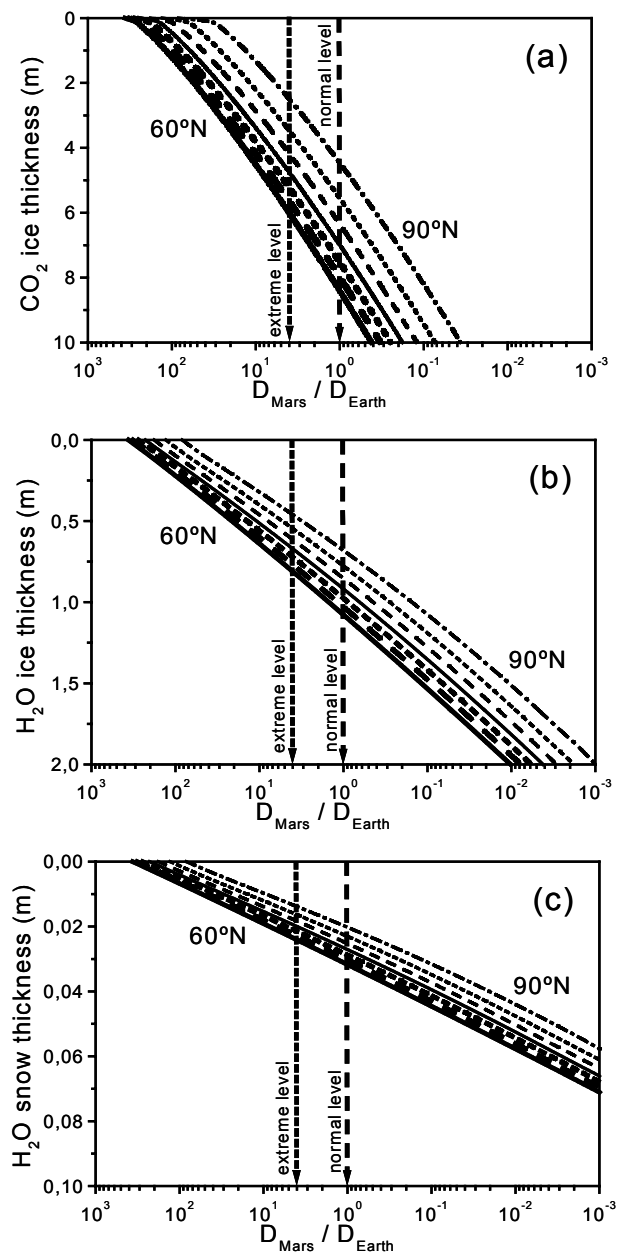

**Fig. 6.** Mapping of the upper and lower ice depth limits (in cm) of the Martian Radiative Habitable Zone (RHZ) for northern polar regions. The upper radiative limit $z_{min}$ at which $D_{Mars} \leq D_{Earth}$ (left column), and thickness ($z_{max}$ - $z_{min}$ ) (right column), where the lower radiative limit $z_{max}$ is such that $PAR_{Mars} \geq PAR_{min}$, are shown for two polar cap scenarios: ($CO_2$ ice + $H_2O$ ice) cover (a and b) and ($CO_2$ ice + $H_2O$ snow) cover (c and d). Seasons for the northern hemisphere are indicated.

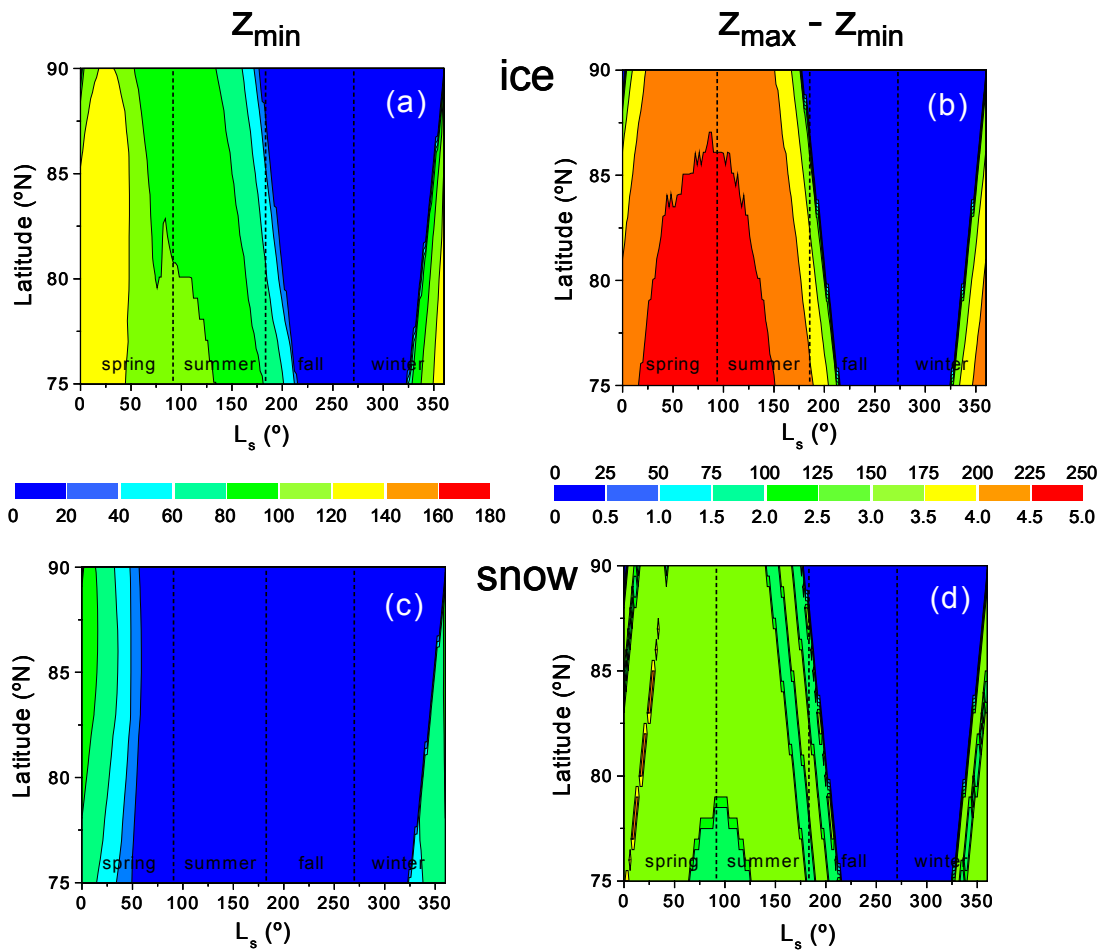

**Fig. 7.** Mapping of the upper and lower ice depth limits (in cm) of the Martian Radiative Habitable Zone (RHZ) for southern polar regions. The upper radiative limit $z_{min}$ at which $D_{Mars} \leq D_{Earth}$ (left column), and thickness ($z_{max}$ - $z_{min}$ ) (right column), where the lower radiative limit $z_{max}$ is such that $PAR_{Mars} \geq PAR_{min}$, are shown for two polar cap scenarios: ($CO_2$ ice + $H_2O$ ice) cover (a and b) and ($CO_2$ ice + $H_2O$ snow) cover (c and d). Seasons for the southern hemisphere are indicated.

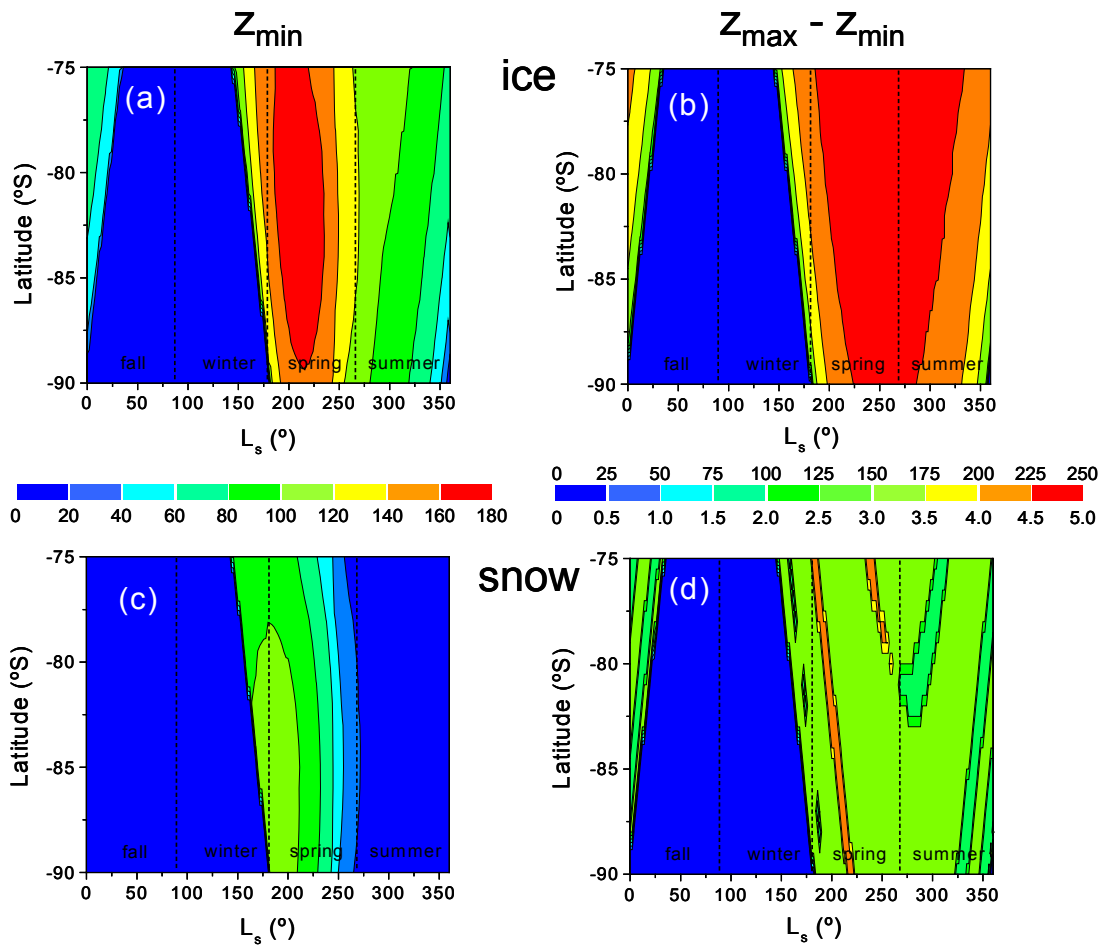

**Fig. 8.** Generalized DNA action spectrum (normalized at 300 nm) used in this work (Setlow 1974; Lindberg and Horneck 1991).

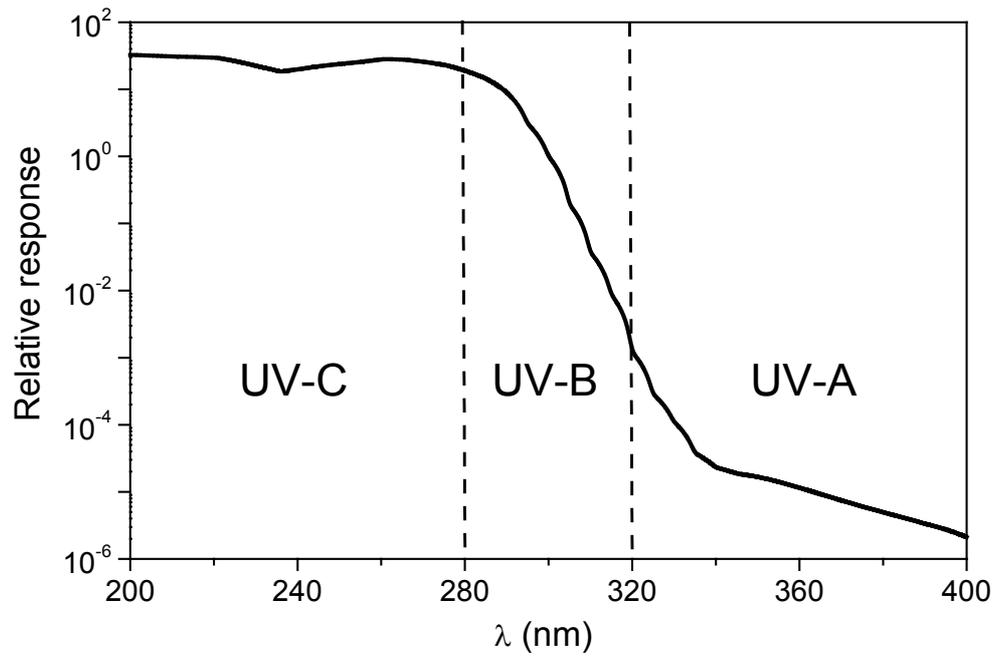